\documentclass[12pt]{article}

\usepackage{amsmath}
\usepackage{amscd}
\usepackage{amssymb}
\usepackage{array}

\begin{document}

\newcommand{\id}{\relax{\rm 1\kern-.28em 1}}

\newcommand{\R}{\mathbb{R}}
\newcommand{\C}{\mathbb{C}}
\newcommand{\Z}{\mathbb{Z}}
\newcommand{\Hb}{\mathbb{H}}

\newcommand{\rSO}{\mathrm{SO}}

\newcommand{\cH}{\mathcal{H}}
\newcommand{\cL}{\mathcal{L}}

\newcommand{\fso}{\mathfrak{so}}
\newcommand{\fk}{\mathfrak{k}}
\newcommand{\fp}{\mathfrak{p}}

\newcommand{\e}{\epsilon}

\rightline{ CERN-PH-TH/2006-044} \rightline{IFIC/06-13}

\vskip 3.5cm

\begin{center}{\LARGE \bf  Real symplectic formulation\\ of local
special geometry}\end{center}

\vskip 1.5cm

\centerline{Sergio Ferrara$^{\flat}$ and \'Oscar Maci\'a$^{\sharp}$}
 \vskip
1.5cm

\centerline{\it $^\flat$
 Physics Department, Theory Unit, CERN,}
\centerline{\small \it CH-1211 Geneva 23, Switzerland }
\centerline{{\footnotesize e-mail: Sergio.Ferrara@cern.ch}}
\medskip
\centerline{\it $^\flat$ INFN, Laboratori Nazionali di Frascati,}
\centerline{\small\it Via Enrico Fermi 40, I-00044 Frascati, Italy}

 \bigskip

 \centerline{\it $^\sharp$
 Departament de F\'{\i}sica Te\`orica,
Universitat de Val\`encia and IFIC}
 \centerline{\small\it C/Dr.
Moliner, 50, E-46100 Burjassot (Val\`encia), Spain.}
 \centerline{{\footnotesize e-mail: Oscar.Macia@ific.uv.es}}


\vskip 1cm

\begin{abstract}
We consider a formulation of local special geometry in terms of
Darboux special coordinates $P^I=(p^i,q_i)$, $I=1,...,2n$. A general
formula for the metric is obtained which is manifestly
$\mathbf{Sp}(2n,\mathbb{R})$ covariant. Unlike the rigid case the
metric is not given by the Hessian of the real function $S(P)$ which
is the Legendre transform of the imaginary part of the holomorphic
prepotential. Rather it is given by an expression that contains $S$,
its Hessian and the conjugate momenta $S_I=\frac{\partial
S}{\partial P^I}$. Only in the one-dimensional case ($n=1$) is the
real (two-dimensional) metric proportional to the Hessian with an
appropriate conformal factor.
\end{abstract}
\vspace{2mm} \vfill
 \vspace{1mm}

\newpage
\section{Introduction}

Local special geometry \cite{WLP}, \cite{S1}, \cite{FS2},
\cite{CFG}, \cite{ABCAFFM}, is the underlying geometry of $N$=2
supergravity in four dimensions  coupled to vector multiplets. It
also emerges in any compactification, obtained from a
higher-dimensional theory, when eight supersymmetries are preserved
in four dimensions or, in a more general setting, when the
lower-dimensional theory can be viewed as a ``deformation'' of  a
$N=2$ locally supersymmetric Lagrangian. These theories include
Calabi--Yau compactifications from \emph{type II} supergravity
\cite{FS2}, \cite{CFG} as well as compactifications on more general
manifolds with $G=SU(3)$ structures \cite{GLMW}, \cite{KSTT},
\cite{CALMZ}, \cite{BBDG}, \cite{GLW}. Special geometry also plays
an important role in the physics of black holes, attractor equations
\cite{FKS}, \cite{FK}, \cite{S2}, \cite{M}, which control the
horizon geometry, and more recently it has been used also  to
connect the black hole entropy to topological partition functions in
superstring
theory \cite{OSV}, \cite{LWKM}, \cite{OVV}.\\

\noindent In this framework it is useful to reconsider a formulation
of special geometry in real \cite{F}, \cite{H}, \cite{ACD}, rather
than complex variables. Real variables, will be denoted as Darboux
coordinates $P^I=(p^i,q_i)$, $i=1,...,n$ where $n$ is the complex
dimension of the (local) special manifold. One of the reasons to
adopt these coordinates is that the attractor equations can be
written as real equations \cite{FKS}, \cite{FK}, \cite{S2},
\cite{M}, \cite{K}, \cite{KSS}, \cite{BFM}, which determine the real
charge vector $Q=(m^\Lambda,e_\Lambda)$, $(\Lambda=0,1,...,n)$ in
terms of the holomorphic sections
$V=(X^\Lambda,F_\Lambda=\partial_\Lambda F)$ of special geometry
\cite{WLP}, \cite{S1}, \cite{FS2}, \cite{ABCAFFM};
 A direct real formulation of the geometry may therefore have a
simplifying role.\\

\noindent In this note we extend a previous investigation \cite{FM}
on the real formulation of rigid special geometry to the local case.
The main result is the derivation of a general formula for the
metric in special Darboux coordinates and also comment on the main
differences from the rigid case.\\

\noindent A very important role is also played here by  a $2n\times
2n$ symplectic real symmetric matrix $M(\Re f_{ij} , \Im f_{ij})$
constructed in terms of the holomorphic matrix
$$f_{ij}=\frac{\partial^2 f(t)}{\partial t^i \partial t^j}$$ where
$t^i=\frac{X^i}{X^0}$ are the ``special coordinates'' of the local
special geometry and $f(t)$ is the holomorphic prepotential in
special coordinates. As in the rigid case the $M$ matrix (which
satisfies $M\Omega M=\Omega$ where $\Omega$ is the invariant
symplectic form of $\mathbf{Sp}(2n,\mathbb{R})$) is related to the
Hessian $$H_{IJ}=\frac{\partial^2 S}{\partial P^I \partial P^J}$$ of
a certain Hamiltonian function $S(p,q)$ written in Darboux special
coordinates but, unlike the rigid case, the general formula for the
metric is given by
\begin{equation}
g_{IJ}(P)=-\frac{1}{2S}H_{IJ}+\frac{1}{4S^2}\left(S_I S_J + (H\Omega
S)_I(H\Omega S)_J\right)
\end{equation}
The first term is analogous to the term that is present in the rigid
case. The second term, due to the Hodge-K\"ahler structure of local
special geometry, has no rigid analogue and depends, other than $S$
and its Hessian, also on the conjugate momenta $S_I=\frac{\partial
S}{\partial P^I}$, of the Darboux coordinates $P^I$. Note that
$(H\Omega S)_ I$ denotes the expression
$$(H\Omega S)_ I=H_{IK}\Omega^{KL}S_L \qquad, \qquad \Omega^{KL}=-\Omega_{KL}$$

\noindent This paper is organized as follows. In section 2 we
describe local special geometry in holomorphic and Darboux special
coordinates and compute the K\"ahler metric in real symplectic
coordinates. In section 3 we compute  the (real symmetric) metric in
the real symplectic Darboux coordinates, and we show the explicit
$\mathbf{Sp}(2n,\mathbb{R})$ covariance  of this metric and its
differences from the rigid case. In section 4 some examples are
given.\\

\section{Local special geometry in Darboux coordinates}
We consider local special geometry in special coordinates
$t^\Lambda=\frac{X^\Lambda}{X^0}=(t^0=1, t^i)$.  In these
coordinates the holomorphic symplectic sections
\begin{equation}
V=(X^\Lambda,\partial_\Lambda F)
\end{equation}
take the simple form \begin{equation} V=\left(1\;,\; t^i \;;\;
f_0=2f-t^if_i\;,\; f_i=\frac{\partial f}{\partial t^i}\right)
\end{equation}
in terms of the prepotential $f(t)=(X^0)^{-2}F(X)$ where $F(X)$ is
homogeneous of degree 2 \cite{WLP} $$X^\Lambda \partial_\Lambda
F=2F$$ From this, it follows that $f(t)$ is a fairly arbitrary
holomorphic function fo $t^i$. Moreover $t^i$ are ``scalar''
sections of the Hodge bundle since under a holomorphic gauge
transformation
\begin{equation}\label{hodgescalar} (X^\Lambda,F_\Lambda)\longmapsto  e^\alpha
(X^\Lambda,F_\Lambda)\quad , \quad \bar\partial_i\alpha=0
\end{equation}
the special coordinates $t^i$ and the prepotential $f(t)$ are invariant.\\

\noindent  Property (\ref{hodgescalar}) is crucial in defining real
coordinates. In fact, while it makes sense to define the real part
of the holomorphic special coordinates $t^i$, it would not make
sense to define a real part of $X^\Lambda$ since this operation
would be incompatible with the Hodge structure of the manifold. Note
that this is also the main difference   between  the local and the
rigid cases \cite{F}, \cite{H}, \cite{M}, \cite{FM}, since in the
latter the Hodge structure is absent and this obstruction does not
arise. We also note that in local special coordinates  the
$\mathbf{Sp}(2n+2,\mathbb{R})$ covariance is lost (by the K\"ahler
gauge-fixing $X^0=1$) but still a $\mathbf{Sp}(2n,\mathbb{R})$
structure is present with respect to the reduced $2n$-dimensional
holomorphic section
\begin{equation}
V_R=\left(t^i, f_i=\frac{\partial f}{\partial t^i} \right)
\end{equation}
In particular, in terms of the matrix $f_{ij}=\frac{\partial^2
f}{\partial t^i \partial t^j}$, we can construct a real symmetric
$2n\times 2n$ symplectic matrix $M(f_{ij})$ as follows \cite{CAF},
\cite{FK}:
\begin{eqnarray}\label{mf}
M(f_{ij})=\left( \begin{array}{cc} \Im f_{ij} + \Re f_{ik} \Im
(f^{-1})^{kl} \Re f_{lj} & -\Re f_{ik}
\Im (f^{-1})^{kj}\\
-\Im (f^{-1})^{ik}\Re f_{kj} & \Im (f^{-1})^{ij}
\end{array} \right)
\end{eqnarray}
with the obvious property $M\Omega M=\Omega$.\\

\noindent We note at this point that the $M(f_{ij})$ matrix
considered here is not the $M(\mathcal{F})$ matrix considered in
\cite{CAF}, \cite{FK}. Indeed while $M(\mathcal{F})\in
\mathbf{Sp}(2n+2,\mathbb{R})$ is a $(2n+2)\times (2n+2)$ symplectic
matrix, the matrix considered here is $2n\times 2n$ and it is
$\mathbf{Sp}(2n,\mathbb{R})$ symplectic. However, it is this last
matrix which plays a role in
our considerations.\\

\noindent The K\"ahler potential of special geometry in these
coordinates is
\begin{eqnarray}
K&=&-log Y\\
\label{y}Y&=&i\left( 2f-2\bar f - (t^i-\bar t^i)(f_i+\bar f_i)
\right)
\end{eqnarray}
from which the following expression for the metric holds:
\begin{equation}\label{premetric}
g_{i\bar j}=\partial_i \bar \partial_j K=-\frac{1}{Y}Y_{i\bar
j}+\frac{1}{Y^2}Y_i Y_{\bar j}
\end{equation}

\noindent By explicit computations the two terms in
(\ref{premetric}) give:
\begin{equation}\label{premetric2}
K_{i\bar j}= -\frac{i}{Y}(f_{ij}-\bar
f_{ij})+\frac{1}{Y^2}\left((f_i-\bar f_i +(\bar
t^k-t^k)f_{ki})\times (\bar f_j -f_j+(t^k-\bar t^k)\bar
f_{kj})\right)
\end{equation}
We now go to Darboux special coordinates exactly as we did in the
rigid case \cite{FM} i.e. we define
\begin{equation}\label{coordinates}
t^i=p^i+i\phi^i \quad, \quad f_i=q_i+i\psi_i
\end{equation}
and the Legendre transform $S(p,q)$ of the imaginary part of the
prepotential
\begin{equation}
L=\Im f
\end{equation}
Then the $(q_i,\phi^i)$ and $(p^i,\psi_i)$ real sections are pairs
of conjugate variables for $L$\begin{equation}\label{legendre}
q_i=\frac{\partial L}{\partial \phi^i} \quad , \quad
\psi_i=\frac{\partial L}{\partial p^i}
\end{equation}
and in terms of $S$ given by
\begin{equation}\label{s}
S(p,q)=q_i\cdot \phi^i(p,q)-L(p,\phi(p,q))
\end{equation}
we have the following set of equations
\begin{equation}
\phi^i=\frac{\partial S}{\partial q_i}\quad , \quad
\psi_i=-\frac{\partial S}{\partial p^i}
\end{equation}
By comparing (\ref{s}) and (\ref{y}) we realize that \cite{OSV},
\cite{OVV}, \cite{GSC}, \cite{LWKM}
\begin{equation} Y=4S
\end{equation} i.e. \begin{equation}\label{expk}S=\frac14 e^{-K}\end{equation}
Eq. (\ref{expk}) is the main difference between rigid and local
special geometry. We just recall that in the former the (square of
the) distance $ds^2$ is given by \cite{F}, \cite{H}, \cite{FM},
\cite{M2}
\begin{eqnarray}\label{rigid}
K_{ij}dz^i\otimes d\bar z^j=-2 H_{IJ}dP^I \otimes dP^J
\end{eqnarray}
where $H_{IJ}=\frac{\partial^2 S}{\partial P^I \partial P^J}$ is the
Hessian of the $S$ functional. In the next section we will see how
the rigid formula (\ref{rigid}) is modified in the local case.\\

\noindent The change of variables from the holomorphic sections
$(t^i, f_i)$ to the Darboux variables $(p^i, q_i)$ and
$(\phi^i,\psi_i)$ is exactly as in the rigid case\; it is still true
\cite{FM} that the $H$ matrix \begin{eqnarray} \label{h} H=\left(
\begin{array}{cc}
\frac{\partial^2 S}{\partial p^i\partial p^j}&\frac{\partial^2
S}{\partial p^i\partial q_j}\\
\frac{\partial^2 S}{\partial q_i\partial p^j} &\frac{\partial^2
S}{\partial q_i \partial q_j}
\end{array} \right)=\left( \begin{array}{cc}S_{ij}& S_i^j\\S^i_j& S^{ij}\end{array}\right)
\end{eqnarray}
is equal to the (negative of the) $M$ matrix defined in (\ref{mf}),
then it follows that $H\Omega H= \Omega$ i.e.
\begin{eqnarray}
S_i^kS_{kj}-S_j^kS_{ki}&=&0\nonumber \\
\label{sss} S^{ik}S_k^j-S^{jk}S_k^i&=&0\\
S_{ik}S^{kj}-S_i^kS_k^j&=&\delta_i^j\nonumber
\end{eqnarray}
By comparing the $M^{ij}$ and $M^i_j$ entries of $M$ with $H$ it
also follows that
\begin{equation}\label{fijss}
f_{ij}=-(S_i^k+i\delta_i^k)(S^{-1})_{kj}
\end{equation}
where $(S^{-1})_{kj}=(S^{kj})^{-1}$ is the inverse of $S^{kj}$
defined by
(\ref{h}).\\

\noindent We can now write Eq. (\ref{premetric2}) in Darboux special
coordinates by means of Eqs. (\ref{coordinates}), (\ref{legendre}),
and (\ref{fijss}) and we finally obtain
\begin{equation}\label{kijs}
K_{i\bar j}=-\frac{1}{2S}(S^{ij})^{-1}+\frac{1}{4S^2}\left(
\frac{\partial S}{\partial p^i}\frac{\partial S}{\partial
p^j}+\frac{\partial S}{\partial q_k}\frac{\partial S}{\partial
q_l}f_{ki}\bar f_{lj}+\frac{\partial S}{\partial q_k}\frac{\partial
S}{\partial p^j}f_{ki}+\frac{\partial S}{\partial q_k}\frac{\partial
S }{\partial p^i}\bar f_{kj}\right)
\end{equation}
where $f_{ij}$ is given by (\ref{fijss}).\\

\noindent We just note that while in the rigid case $K_{i\bar
j}=-2(S^{ij})^{-1}$, in the local case (\ref{kijs}) also depends on
$S$ and the canonical momenta $S_I=\frac{\partial S}{\partial P^I}$.
The second term in (\ref{kijs}) is the main difference between the
rigid and local case.\\

\section{The real symplectic metric}

In order to compute the real metric we have to transform the
``distance'' from holomorphic to Darboux special coordinates
\begin{equation}\label{comparison}
ds^2=K_{i\bar j}dt^i\otimes d\bar t^j=g_{IJ}dP^I\otimes dP^J
\end{equation}
The result is obtained by computing the differentials $dt^i\otimes
dt^j$ in Darboux coordinates and then comparing the two expressions
in Eq. (\ref{comparison}).\\ We just mention that the computation of
$dt^i\otimes d\bar t^j$ is exactly the same as in the rigid case
\cite{FM} since the Legendre transform and the conjugate variables
are the same. Indeed only Eqs. (\ref{h}), (\ref{sss}), (\ref{fijss})
which are the same as in rigid special
geometry really matter.\\

\noindent The change of variables in the differentials was given in
Eq. (50) of reference \cite{FM} and it is reproduced here for the
benefit of the reader.
\begin{eqnarray}\label{reproduction}
dt^i&\otimes& d\bar t^j =
\left(\delta^i_k\delta^j_l+\frac{\partial^2 S }{\partial
q_i\partial p^k}\frac{\partial^2 S }{\partial q_j\partial
p^l}+i\left(\frac{\partial^2 S }{\partial q_i\partial
p^k}\delta^j_l-\frac{\partial^2 S }{\partial q_j\partial
p^l}\delta^i_k \right)\right)dp^k\otimes dp^l \nonumber
\\&+& \left( \frac{\partial^2 S }{\partial q_i\partial
q_l}\frac{\partial^2 S }{\partial q_j\partial
p_k}+\frac{\partial^2 S }{\partial q_i
\partial p_k}\frac{\partial^2 S }{\partial q_j \partial
q_l}+i\left(\frac{\partial^2 S }{\partial q_i \partial
q_l}\delta^j_k-\frac{\partial^2 S }{\partial q_j\partial
q_l}\delta^i_k\right)
 \right)dp^k\otimes
dq_l\nonumber
\\&+&\left(\frac{\partial^2 S }{\partial q_i \partial
q_k}\frac{\partial^2 S }{\partial q_j \partial q_l}\right)dq_k
\otimes dq_l
\end{eqnarray}

\noindent By explicit multiplication of (\ref{kijs}) with
(\ref{reproduction}) we finally obtain, from the first term in
(\ref{kijs})
\begin{equation}
-\frac{1}{2S}(S^{ij})^{-1}dt^i\otimes d\bar t^j=
-\frac{1}{2S}H_{IJ}dP^I \otimes dP^J
\end{equation}
For the second term of (\ref{kijs}) (by multiplying by $dt^i\otimes
d\bar t^j$ ) we find
\begin{equation}\label{secondterm}
-\frac{1}{8S^2}\left(S^KH_{KI}S_J-S^KH_{KJ}S_I
\right)\left(dP_LH^{LI}dP^J-dP_LH^{LJ}dP^I \right)
\end{equation}
where $$S^K=\Omega^{KI}S_I \quad , \quad H^{LJ}=(H_{LJ})^{-1}
\quad ,\quad dP_L=\Omega_{LI}dP^I$$ and
$$\Omega^{KI}=(\Omega_{KI})^{-1}=-\Omega_{KI}$$
\noindent  By multiplying the four terms in (\ref{secondterm}) we
finally obtain:
\begin{equation}
\frac{1}{4S^2}\left(S_IS_J+(H\Omega S)_I (H\Omega  S)_J  \right)
\end{equation}
where $$(H\Omega S)_I=H_{IK}\Omega^{KL}S_L$$ The final expression
for the real symmetric metric in Darboux coordinates, for local
special geometry, is therefore given by
\begin{equation}\label{secondterm2}
g_{IJ}(P)=-\frac{1}{2S}H_{IJ}+\frac{1}{4S^2}\left(S_I S_J + (H\Omega
S)_I(H\Omega S)_J\right)
\end{equation}
We note at this point that a major simplification occurs for $I=1,2$
which corresponds to one-dimensional complex special geometry.
Indeed in this case
\begin{eqnarray}\label{simplification}
S^KH_{KI}S_J-S^KH_{KJ}S_I=-\Omega_{IJ}S_K(H^{-1})^{KL}S_L\\
dP_LH^{LI}dP^J-dP_LH^{LJ}dP^I=-\Omega^{IJ}dP^LH_{LK}dP^K
\end{eqnarray}
and then (\ref{secondterm}) becomes
\begin{equation}
\frac{1}{4S^2}S_K(H^{-1})^{KL}S_LH_{IJ}dP^I\otimes dP^J
\end{equation}
In this case, the metric therefore is \begin{equation}\label{corta}
g_{IJ}=\left(-\frac{1}{2S}+\frac{1}{4S^2}(SH^{-1}S)\right)H_{IJ}
\end{equation}
We finally note that the metric given by (\ref{secondterm2}) has a
manifest $\mathbf{Sp}(2n,\mathbb{R})$ covariant structure. The
simplification occurring in (\ref{simplification}) is due to the
fact that for  $\mathbf{Sp}(2,\mathbb{R})$ the antisymmetric
representation is the identity while the traceless part is
non-vanishing for $n>1$; and this is the reason why
(\ref{secondterm2}) and (\ref{corta}) are equivalent only for
$n=1$.\\

\noindent It is obvious that in special coordinates the
$\mathbf{Sp}(2n+2,\mathbb{R})$ structure of the geometry is lost and
only a $\mathbf{Sp}(2n,\mathbb{R})$ structure is manifest.
Nevertheless we have shown that this is what is needed to obtain a
general expression for the metric  in terms of the Darboux data of
local special geometry. The $\mathbf{Sp}(2n+2,\mathbb{R})$
covariance will be considered elsewhere.

\section{Some examples}
In this section we compute the real metric in some examples of
special geometry. The two examples in question correspond to a cubic
and quadratic prepotential, for one complex scalar:
\begin{eqnarray}
f(t)&=&\frac13t^3\\
\label{secondex}f(t)&=&\frac{i}{4}(t^2-1)
\end{eqnarray}

\noindent Let us first consider the cubic case. The $S$ functional
is
\begin{equation}\label{4sy}
4S=Y=-\frac83(\Im t)^3
\end{equation}
where $S=-\frac23(\Im t)^3$ so that $\Im t < 0$. Note that in this
case $\Im f_{tt}=2\Im t<0$ and therefore the Hessian will be
positive definite. Going to Darboux coordinates we find
\begin{equation}\label{simple}
q=p^2-\phi^2 \longrightarrow \phi=\pm (p^2-q)^{\frac12}
\end{equation}
so we have $p^2>q$ and also the negative root must be chosen such
that $S$ is positive. From (\ref{simple}) and (\ref{4sy}) we obtain
\begin{equation}
S=\frac23(p^2-q)^{\frac32}
\end{equation}

\noindent The Hessian matrix is given by
\begin{equation}
H=(p^2-q)^{-\frac12}\left(\begin{array}{cc} 2(2p^2-q)&-p\\
-p& \frac12 \end{array}\right)
\end{equation}
Since $Det(H) =1$ and also $(p^2-q)^\frac12 Tr(H)=2(2p^2-q)+\frac12>
0$, $H$ is indeed positive definite.

\noindent We can now compute the metric by using (\ref{corta}).
First, by computing the second term, we obtain
\begin{equation}
SH^{-1}S= 2(p^2-q)^{\frac32}
\end{equation}
and then \begin{equation}
-\frac{1}{2S}+\frac{SH^{-1}S}{4S^2}=\left(-\frac34+
\frac98\right)(p^2-q)^{-\frac32}=\frac38(p^2-q)^{-\frac32}
\end{equation}
so that the metric is
\begin{equation}
g_{IJ}(p,q)=\frac38(p^2-q)^{-2}\left(
\begin{array}{cc}2(2p^2-q)&-p\\
-p&\frac12\end{array}\right)
\end{equation}

\noindent We now turn to the second example (\ref{secondex}) where
now
\begin{equation}
Y=4S=1-t\bar t = 1-p^2-\phi^2
\end{equation} therefore we must take $p^2+\phi^2<1$. Since $$\Im
f=\frac14(p^2-\phi^2-1)=L$$ we have $$q=\frac{\partial L}{\partial
\phi}=-\frac12 \phi \qquad \longrightarrow \qquad \phi=-2q$$ It
follows that
\begin{equation}
Y=4S=1-p^2-4q^2 \quad , \quad S=\frac14-\frac{p^2}{4}-q^2
\end{equation}

\noindent The Hessian matrix is
\begin{equation}
H=\left(\begin{array}{cc}-\frac12 & 0\\ 0 & -2\end{array} \right)
\end{equation} and it is negative definite. The conjugate momenta
are $$\frac{\partial S}{\partial p}=-\frac{1}{2}p \quad , \quad
\frac{\partial S}{\partial q}=-2q$$ We then have
$$SH^{-1}S=-\frac12(p^2+4q^2)$$ and the conformal factor becomes
$$-2\frac{1}{(1-p^2-4q^2)^2}$$ The local metric is then finally given by
\begin{equation}
g_{IJ}(p,q)=\frac{1}{(1-p^2-4q^2)^2}\left( \begin{array}{cc} 1 & 0\\
0 & 4
\end{array}\right)
\end{equation}

\section{Acknowledgements}We would like to thank M.A. Lled\'o for enlightening
discussions.\\
S.F. would like to thank the Theoretical Physics Department at the
University of Valencia where part of this work was performed. S.F.
was partially supported by funds of the INFN-CICYT bilateral
agreement. The work of S.F. has also been supported in part by the
European Community Human Potential Program under contract
MRTN-CT-2004-005104 ``Constituents, fundamental forces and
symmetries of the Universe'', in association with INFN Frascati
National
Laboratories  and by the D.O.E. grant DE-FG03-91ER40662, Task C.\\
O.M. wants to thank the Department of Physics, Theory Division at
CERN for its kind hospitality during the realization of this work.
The work of O.M. has been supported by an FPI fellowship from the
Spanish Ministerio de Educaci\'on y Ciencia (MEC), through the grant
FIS2005-02761 (MEC) and EU FEDER funds, by the Generalitat
Valenciana, contracts GV04B-226, GV05/102 and by the EU network
MRTN-CT-2004-005104 ``Constituents, fundamental forces and
symmetries of the Universe''.

\end{document}